\documentclass{edp-jp4}
\usepackage{graphicx}
\usepackage{amsmath}
\usepackage{amssymb}

\begin{document}

%%-----------------------------
%%      the top matter
%%-----------------------------
\title{One-dimensional behavior of elongated Bose-Einstein condensates}
\author{P.\ Bouyer}\address{Laboratoire Charles Fabry de l'Institut d'Optique, UMR 8501 du CNRS, 91403 Orsay Cedex, France}
\author{J.\ H.\ Thywissen}\address{Department of Physics, 60 St George Street, University of Toronto, Toronto, ON, M5S 1A7, Canada}
\author{F.\ Gerbier}\sameaddress{1}
\author{M.\ Hugbart}\sameaddress{1}
\author{S.\ Richard}\sameaddress{1}
\author{J. Retter}\sameaddress{1}
\author{A. Aspect}\sameaddress{1}
\maketitle
\begin{abstract} We study the properties of elongated Bose-Einstein
condensates. First, we show that the dimensions of the condensate after expansion differs from the 3D Thomas-Fermi regime. We also study the coherence length of such elongated condensates. \end{abstract}
%
%%-----------------------------
%%      your text
%%-----------------------------
\section{Introduction}
\label{intro}
The experimental study of one dimensional (1D) degenerate Bose
gases, where the radial motion is``frozen'', is currently an
important direction of research in ultra-cold atom physics
\cite{gorlitz01,schreck01,greiner01}. In uniform 1D systems,
fluctuations of the condensate phase are
pronounced, because of a large population of low-lying states; as
a result, no long range order exists at any temperature. In a
trapped 1D gas, the finite size of the sample naturally introduces
a low-energy cutoff, and at sufficiently low temperature a phase
coherent sample can exist (see \cite{petrov1d} and references
therein for the properties of 1D gases). Essentially, the same
analysis holds for three dimensional (3D) trapped gases in very
elongated traps \cite{petrov3d}: here the interaction energy per
particle is larger than the radial frequency, and the condensate
wavefunction is built from several radial modes. Still, low-energy
excitations with a frequency smaller than the radial frequency are
effectively one-dimensional, and induce large fluctuations of the
condensate phase \cite{petrov3d}. Such condensates with a
fluctuating phase, or {\it{quasicondensates}}, have been
observed experimentally in \cite{dettmer01,hellweg01,shvarchuck02,Richard03,Ertmer03}.

We present here an overview of our experimental study of elongated condensates in a very
anisotropic trap. Our set-up, presented in section \ref{creation} allows us to characterize precisely the properties of degenerate gases between the 3D and 1D regimes. In this regime, we first show (section \ref{exp}) that the (quasi)condensates exhibit the same smooth profile as a true (fully coherent)
condensate. This profile, however, is affected by both the radial quantum pressure (zero point oscillation in the tight trapping potential) and the presence of an interacting thermal cloud surrounding the condensate and interacting with it through two-body repulsive interactions. In section \ref{fluctuations}, we  discuss the coherence length of such elongated condensates. Using Bragg spectroscopy \cite{stenger99,Richard03,Fabrice_theory,Gerbier}, we show in section \ref{bragg} that the coherence length is indeed strongly affected by large fluctuations of the
condensate phase \cite{petrov3d} and that
 the measured momentum width depends strongly on temperature; the corresponding coherence length
$L_\phi$, in the ratio $5<\frac{L}{L_\phi}<30$, is in good agreement with theory \cite{petrov3d,Fabrice_theory}.  For larger coherence length, Bragg spectroscopy becomes
increasingly difficult to apply,  and a second method, using a $\pi/2-\pi/2$ interferometer similar to \cite{Ertmer03} is presented in section \ref{interf}.

\section{Creation of elongated BEC in a Iron-Core Electromagnet}
\label{creation}

The experimental setup has been described elsewhere
\cite{interrupted}. Briefly, a Zeeman-slowed atomic beam of
$^{87}$Rb is trapped in a MOT, and after optical pumping into the
$F=1$ state is transferred to a magnetic Ioffe-Pritchard trap
created by an iron-core electromagnet. A new design allows us to
lower the bias field to a few Gauss and thus to obtain very tight
radial confinement \cite{these_vincent}. Final radial and axial
trap frequencies ($\omega_\perp$ and $\omega_z$) are respectively
760 Hz and 5 Hz (i.e. aspect ratio = 152). In this way, we obtain needle-shaped condensates
containing around $5 \times 10^4$ atoms
\cite{calib_N0}, with a typical half-length
$L\simeq130\,\mu$m and radius $R\simeq0.8\,\mu$m. Since the chemical potentiel $\frac{\mu_{\rm TF}}{\hbar \omega_\perp}\sim 4$ (see definition in section \ref{exp}), 
the clouds are
between the 3D and 1D Thomas-Fermi (TF) regimes \cite{gorlitz01,schreck01,greiner01}.
The measurement of the number of atoms is calibrated from a measurement of $T_C$. The temperature $T$ is extracted from a two-component fit to the absorption images or is extrapolated
from the final frequency of the rf ramp when the thermal fraction
is indiscernible.
After the last
evaporation ramp, we observe the formation of a condensate that
exhibits strong shape oscillations (see \cite{DalfovoRMP} and
references therein). This is in line
with a recent observation by Shvarchuck {\it{et al.}}
\cite{shvarchuck02}, who propose the picture of a BEC locally in
equilibrium at each point on the axis, but globally out of
equilibrium. Indeed, the radial degrees of freedom are rapidly
thermalized, but axial motion damps on a much longer time scale.
We therefore hold the condensate for a variable time (typically 7
seconds) in presence of an rf shield, so that these collective
oscillations have no influence on the Bragg spectra.

\section{Expansion of elongated condensates}
\label{exp}

In anisotropic traps, the aspect ratio of the ballistically expanding cloud can be used to find a signature of  Bose-Einstein Condensation. For a Bose gas above critical temperature, the expanding cloud is expected to become isotropic at long expansion times (typically for times larger than $1/\omega_z$, where $\omega_z$ is the smaller trap oscillation frequency). On the contrary, a Bose-Einstein condensate of initial aspect ratio $\epsilon=\omega_z/\omega_\perp$ will only become isotropic at a fixed time $\sim\omega_{z}^{-1}$  and then will invert its anisotropy for longer expansion time. 

The transition from a 3D to a lower dimension (1D) condensate is also observable through the expansion of the degenerate gas \cite{gorlitz01}. For a Bose-Einstein condensate in the 3D Thomas-Fermi regime ($\mu \gg \hbar\omega_\perp$), the half-length  $L_{\rm TF}$(long axis) and the radius  $R_{\rm TF}$ (confined dimension) of the condensate vary similarly with the number of atoms $N_0$. More precisely, one finds the
3D Thomas-Fermi length $L_{\rm TF}^2=2\mu/(M\omega_z^2)$ and
radius $R_{\rm TF}=\epsilon L_{\rm TF} $, with $\mu$ given by
$2\mu=\hbar\bar{\omega}(15N_0a/\sigma)^{2/5}$ \cite{DalfovoRMP},
where $\bar{\omega}=(\omega_\perp^2\omega_z)^{1/3}$,
$\sigma=\sqrt{\hbar/(M\bar{\omega})}$ and $a=5.32$\,nm
\cite{scat_length}. As a result, the initial degree of anisotropy
$\epsilon=R/L=\omega_z/\omega_{\perp}\ll1$ is independent of
the number of atoms. After a period of free expansion of duration
$t$, the dimension of the expanding condensate are found using the
scaling solutions \cite{castin_kagan}:
\begin{equation}
R(t) = R_{\rm TF}\sqrt{1+ \tau^2}.
\label{expansion_R}
\end{equation}

\begin{equation}
L(t) = L_{\rm TF}(1+\epsilon^2 \left[ \tau \arctan\tau-\ln\sqrt{1+\tau^2} \right])\simeq L_{\rm TF}.
\label{expansion_L}
\end{equation}
with the dimensionless time variable $\tau = \omega_\perp t$. Equations (\ref{expansion_R},\ref{expansion_L}) show that the aspect ratio is independent of
the atom number at any time, and approaches $\omega_{\rm } t$ for
$\tau \gg 1$

As one pushes towards the 1D regime ($\mu\sim\hbar\omega_\perp$),
the radial expansion is increasingly affected by radial zero-point
motion in the trap, neglected in the 3D TF approximation.
Ultimately, the radial size in the trap should approach
$a_\perp=\sqrt{\hbar/M\omega_\perp}$, and the radial expansion
velocity $\omega_\perp a_\perp=\sqrt{\hbar\omega_\perp/M}$. Since
axially the TF approximation ($\mu \gg \hbar\omega_z$) is still valid, the length still
depends on $N_0$, and so does the aspect ratio. Accordingly, a
dependence of the aspect ratio on $N_0$ at near-zero temperature can
be seen as a signature of the breakdown of the TF approximation in
the radial direction \cite{gorlitz01}.

\begin{figure}[!t]
\hspace{-.25cm}
\includegraphics[width=4.5cm]{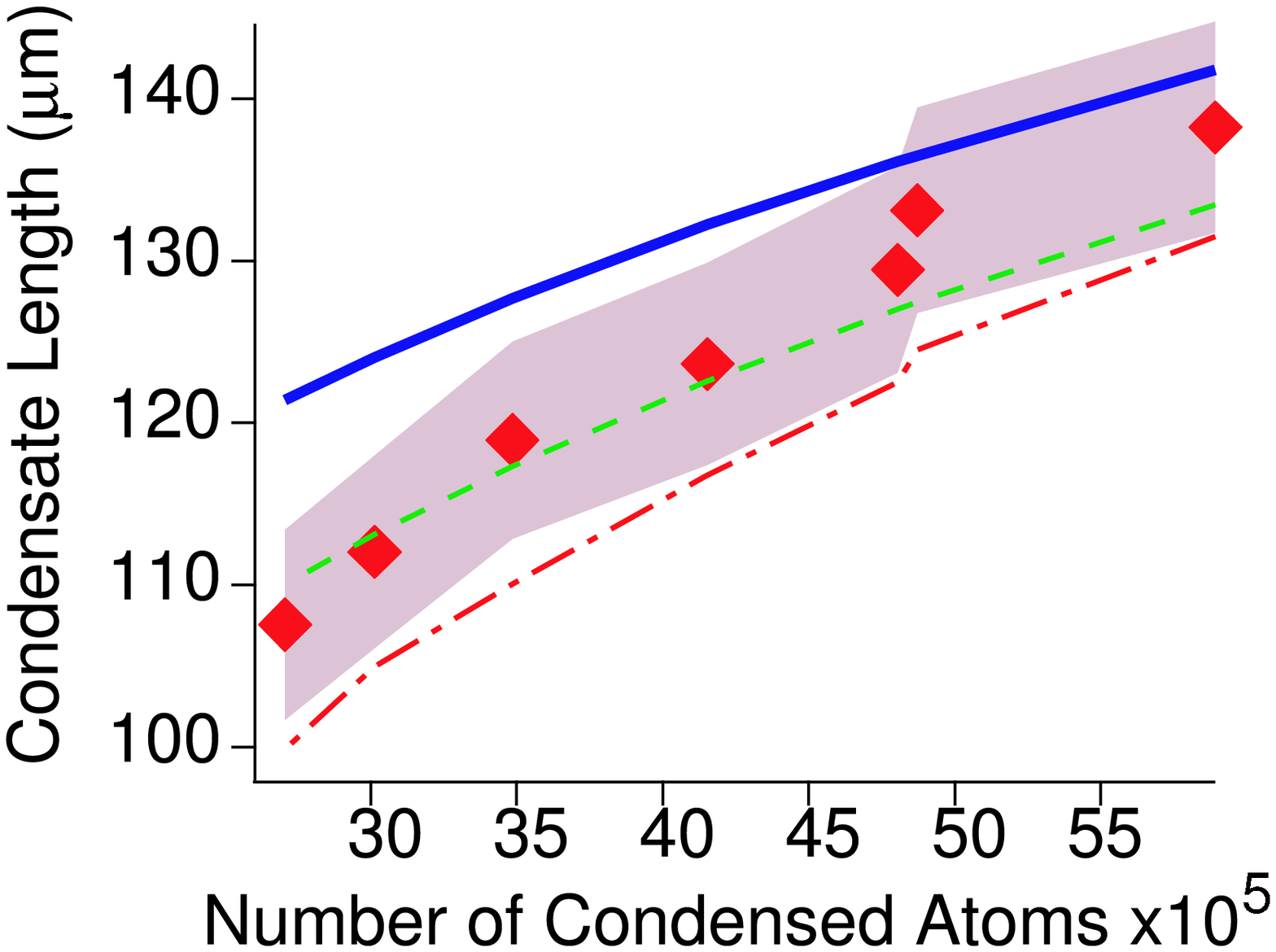}
\hspace{-4.5cm}\vskip-2.5cm\includegraphics[width=4.5cm]{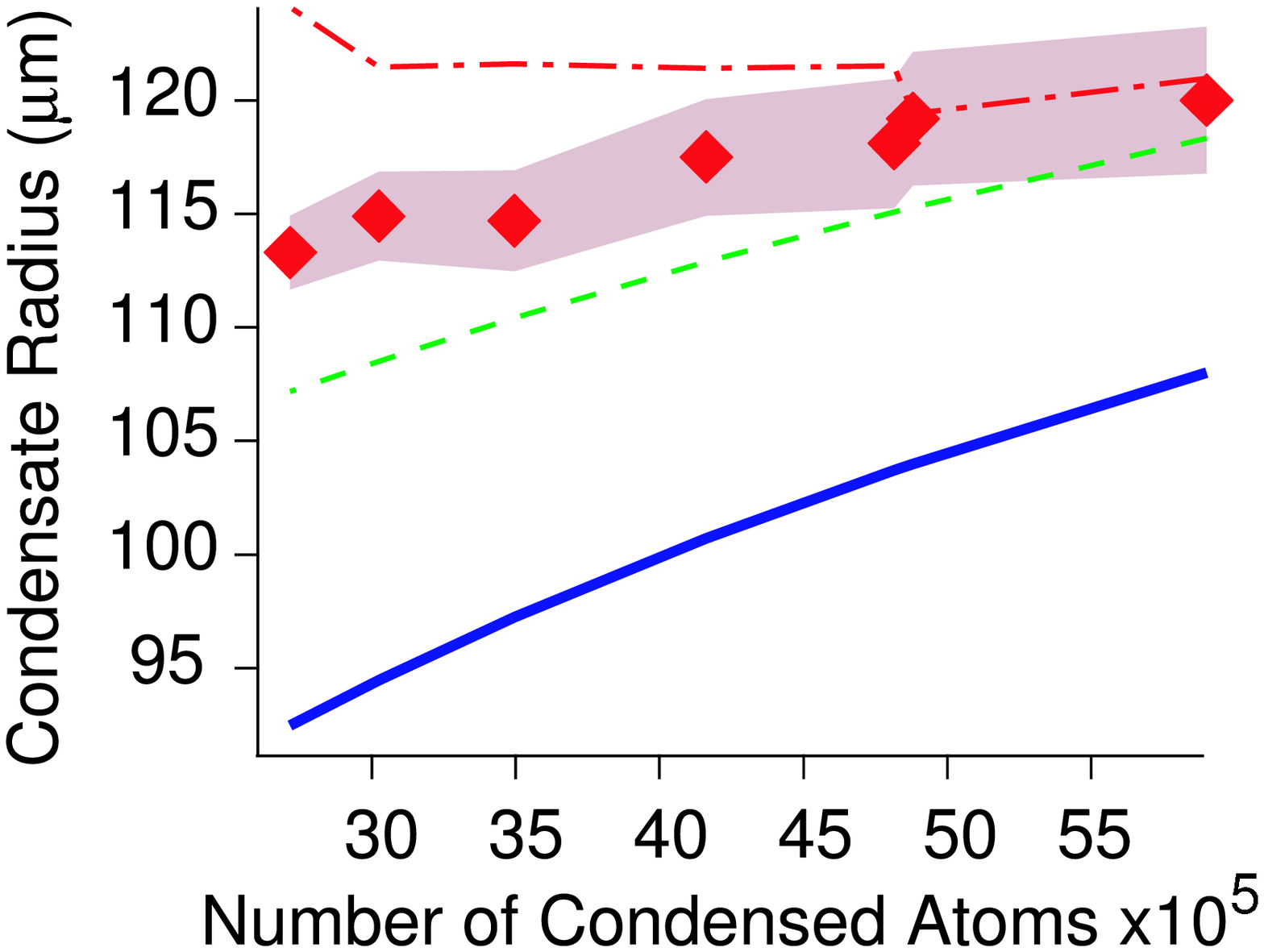}
\includegraphics[width=7.5cm]{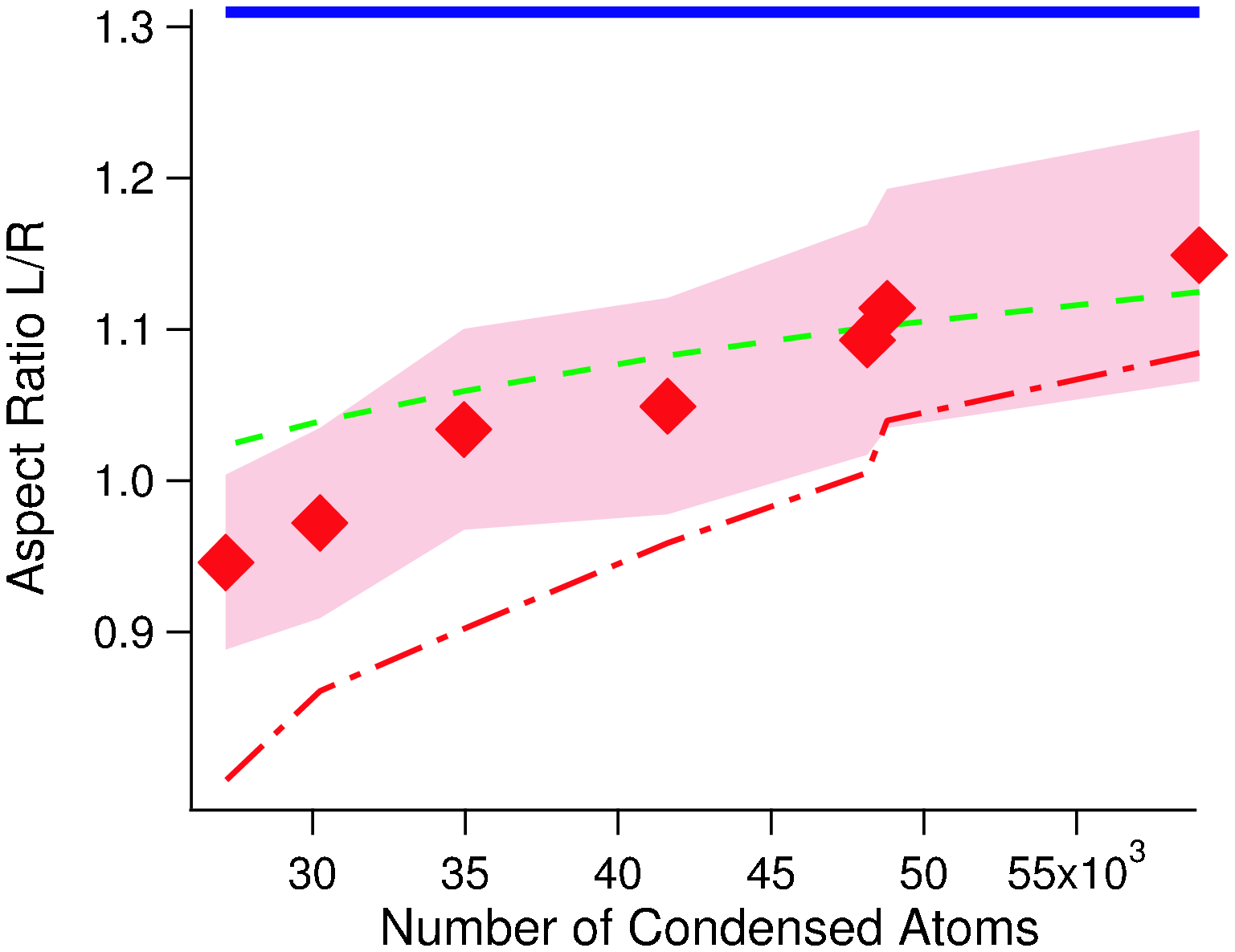}
\caption{Half-length $L$ (upper-left), radius $R$ (lower-left) and aspect ration (right) of the quasi-1D condensate versus
the condensed atom number after a 24.3 ms time of flight. Experimental values (diamonds) are compared to the standard $T=0$
Thomas-Fermi (plain),  and the self consistent solution of equations (\ref{eq:L}) and
(\ref{zubiref}) (dashed). The effect of quantum pressure alone is also shown (dot-dashed).
Calculations use the measured temperature and number of condensed
atoms, as determined by a two-component fit to averaged absorption
images. Combined statistical and systematic errors are indicated with the confidence band (grayed area) around the experimental points.}
\label{longueur_vs_tsurtphi}
\end{figure}

To show how quantum pressure manifests itself in our experiment, 
we first turn to the measurement of the axial half-length $L$ after
release, which faithfully reflects the half-length in the trap since
axial expansion is negligible (eq. \ref{expansion_L}) for our experimental parameter.
Figure~\ref{longueur_vs_tsurtphi} shows that the measured values
of $L$ for various number of condensed atoms (filled diamonds) are smaller than
the standard Thomas-Fermi prediction (solid line).%
This smaller value results from radial quantum pressure (1D), and to a slightly larger extent from
the compression of the quasi-condensate by the 3D thermal
component (excitations with energy much larger than
$\hbar\omega_\perp$). 

We take into account radial quantum
pressure in this crossover regime by finding the optimal ground state energy \cite{Zubarev_1D}
\begin{equation}
\frac{E}{N}=\max_{\gamma_x,\gamma_y,\gamma_z}[\sum_{i=1}^3\frac{\hbar \omega_i}{2}\sqrt{\gamma_i}+\frac{5}{14}(\frac{15}{4\pi} gN m^{3/2}\prod_{i=1}^3(\sqrt{1-\gamma_i}\omega_i))^{2/5}].
\label{zubiref}
\end{equation}
where $\gamma_i$ are effective parameters taking the
zero-point energy contribution in the direction $i=\{x,y,z\}$ into
account. Solving (eq. \ref{zubiref}) allows then to calculate a modified aspect ratio in the trap
$\tilde{\epsilon}=\gamma_z/\gamma_{x,y}$. To include the effect of the thermal cloud, we use the Hartree-Fock approach \cite{intmodel,DalfovoRMP} and calculate the modified length :
{\small\begin{equation}
 L^2=\frac{2g}{M\omega_z^2}\left\{n_0(0)+\frac{2}{\lambda_{\rm T}^{3}}
 \left[{\textsl g}_{3/2}(e^{-\frac{gn_0(0)}{k_{\rm B}T}})-{\textsl
 g}_{3/2}\left( 1 \right)\right]\right\},
\label{eq:L}\end{equation}}with the coupling constant
$g=4\pi\hbar^2a/M$, the thermal de Broglie wavelength
$\lambda_{\rm T}=[2\pi\hbar^2/(M k_{\rm B} T)]^{1/2}$, and
${\textsl g}_{3/2}(x)=\sum _{n=1}^{\infty}x^n/n^{3/2}$. The black dashed line in Fig.~\ref{longueur_vs_tsurtphi} shows the solution of
Eq.~(\ref{eq:L}) assuming a parabolic profile, such that
$n_0(0)=15N_0 L^{-3}\tilde{\epsilon}^2/(8\pi)$. The modified aspect ratio in the trap
$\tilde{\epsilon}$ is calculated according to \ref{zubiref}
and by a self-consistent iteration of equations (\ref{eq:L}) and
(\ref{zubiref}) we obtain $L$ and $\gamma_{x,y,z}$. We find that the calculated lengths are close to our
measurements to within our estimated calibration uncertainty (shown by the grayed area explained in the caption).

We then turn to the measurement of the cloud radius after 24.3 ms of time of flight.
Figure~\ref{longueur_vs_tsurtphi} shows that the measured values
of $R$ at various condensed atom number (filled diamonds) are larger by about $20 \%$ than
the standard Thomas-Fermi prediction (filled line). This increased radius results again from quantum pressure, which leads to an excess released energy in the transverse direction. Using again \cite{Zubarev_1D}, the radius after time of flight is found to be:
\begin{equation}
R(t)^2=R^2_{\rm\small TF}\frac{\prod_{i=1}^3({1-\gamma_i})^{1 \over 5}}{\sqrt{1-\gamma_{x,y}}}(1+\tau^2),
\label{rayzu}
\end{equation}
with $\gamma_i$ calculated from  (eq. \ref{zubiref}).
Calculating the effect of the compression of the thermal cloud and
the subsequent evolution in time of flight is more complex, since it involves energy exchange between both components.  By making the extreme assumption that all the interaction energy is converted to kinetic energy into the condensate after
expansion, we can use again equation \ref{rayzu} with correction from \ref{eq:L} and
\ref{zubiref}. The results presented in figure~\ref{longueur_vs_tsurtphi} show that, both approaches agree with the experiment to within our calibration errors. 

Finally, we observe a clear dependence of the aspect ratio on the condensed atom number, which can be correctly reproduced by our theoretical approach. This proves that in our trap, with this atom number ($\sim 5\times10^4$), the condensate is at the onset of 1D.

\section{Measurement  of the 1D phase fluctuations}
\label{fluctuations}

Phase coherence of the (quasi)condensate
can be characterized by the $1/e$ decay
length of the spatial correlation function, $\mathcal{C}^{(1)}(s)=1/N
\int d^{3}{\mathbf{R}} \langle \hat \Psi^{\dagger}({\mathbf{R}}+s
{\mathbf{e_{\mathrm{z}}}}/2) \hat\Psi( {\mathbf{R}}-s
\mathbf{e_\mathrm{z}}/2) \rangle$ \cite{zambelli00}, with $\hat\Psi$ the
atomic field operator and ${\mathbf{e_{\mathrm{z}}}}$ the unit vector in the axial direction.
Consider a sample of $N_{0}$ condensed atoms, trapped in a
cylindrically symmetric, harmonic trap with an aspect ratio
$\lambda=\omega_{\mathrm{z}}/\omega_{\perp} \ll 1$, in the 3D
Thomas-Fermi regime.
Petrov {\em et al.} \cite{petrov1d,petrov3d} investigated the long wavelength
fluctuations of the phase of $\Psi$, with the following conclusions.
Below a characteristic temperature $T_{\phi}=15 N_{0} (\hbar
\omega_{\mathrm{z}})^{2}/32\mu$, the phase fluctuations are small,
and the coherence length is essentially the condensate size $L$,
as in ``ordinary'' 3D condensates \cite{stenger99,hagley99} (to
put it differently, $\mathcal{C}^{(1)}$ is limited by the density
envelope of $\Psi$). On the other hand, for $T \gg T_{\phi}$ the
coherence length is limited to a value much smaller than $L$, the
value at the center of the trap being $L_{\phi}= L T_{\phi}/T$.

The theoretical results obtained in \cite{petrov3d,
Fabrice_theory} do not apply directly to our experiment. In those
works, a 3D Thomas-Fermi density profile was assumed, whereas in
our case modification of the density profile by the thermal cloud
and radial quantum pressure must be accounted for. As discussed in the previous section, we indeed find 
that a parabolic profile is still a good fit
function, but that the usual $T=0$ relations between $\mu$, $L$,
$R$, and the number of condensed atoms $N_0$ are no longer valid.
We therefore extend the calculation of the axial correlation
function in \cite{Fabrice_theory} to an arbitrary density profile,
in the local density approximation. In the mean field regime
\cite{petrov1d}, from the result of \cite{kane_kadanoff} for a 1D
uniform Bose gas at finite temperature, we obtain: {\small
\begin{equation}C(s)=\int dz\, n_{1}(z)\,{\rm
exp}(-\frac{n_{1}(0)|s|}{2n_{1}(z)L_\phi}),\label{equ:C}
\end{equation}}where $n_{1}(z)=\int d^2{{\bf r}_\perp}\,n_0({\bf r}_\perp,z)$ is
the axial 1D density of the quasi-condensate, while $n_0({\bf r})$
is its 3D density profile. The coherence length near the center of
the trap is given by $L_\phi=\hbar^2 n_{1}(0)/(Mk_{\rm B}T)$
\cite{TFlimits}. Following Petrov {\it et al.}
\cite{petrov1d,petrov3d}, we define the temperature which
delineates the border between coherent and phase-fluctuating
condensates as $T_{\phi}=L_\phi T /L$. Since $n_1(0)$, $L$, and
$T$ are extracted directly from the images, the definitions of
$L_\phi$ and $T_\phi$ relate the coherence properties to
experimentally measured quantities. The axial momentum
distribution follows from a Fourier transform of $C(s)$ and is
well approximated by a Lorentzian of width $\Delta p_{\phi} =
\alpha \hbar/L_\phi$ (HWHM), with $\alpha=0.67$ for a 3D Thomas-Fermi condensate \cite{Fabrice_theory,alpha}. 

\subsection{Momentum spectroscopy with Bragg diffraction}
\label{bragg}

A convenient technique to measure the coherence length is to 
analyse the momentum distribution along the long direction of the condensate.
Our measurement is based on four-photon
velocity-selective Bragg diffraction. Atoms are diffracted out of
the condensate by interaction with a moving standing wave, formed
by two counter-propagating laser beams with a relative detuning
$\delta$ \cite{stenger99,davidson}. Due to the Doppler effect, the
momentum component resonantly diffracted out of the condensate is
$p_z=M(\delta-8\omega_{\rm{R}})/(2k_{\rm L})$ with
$\omega_{\rm{R}}=\hbar k_{\rm{L}}^2/(2M)$, $M$ the atomic mass,
and $k_{\rm L}=2\pi/\lambda$ ($\lambda=780.2$\,nm). The predicted
spectral width is therefore $\alpha \Delta\nu_\phi$, where {\small
\begin{equation} \Delta \nu_{\phi}= \frac{2\hbar k_{\rm L}} {2 \pi
M L_{\phi}}.
\end{equation}}

To build the momentum spectrum of the quasi-condensate, we measure
the fraction of diffracted atoms versus the detuning $\delta$
between the counter-propagating laser beams. The differential
frequency $\delta$ must be stable to better than the desired
spectral resolution, about 200\,Hz for our typical
$L_\phi=10\,\mu$m. The optical setup is as follows. A single laser
beam is spatially filtered by a fiber optic, separated into two
arms with orthogonal polarizations, frequency shifted by two
independent 80\,MHz acousto-optic modulators, and recombined. The
modulators are driven by two synthesizers stable to better than
1\,Hz over the typical acquisition time of a spectrum. The
overlapping, recombined beams are then sent through the vacuum
cell, parallel (to within 1\,mrad) to the long axis of the trap,
and retro-reflected to obtain two standing waves with orthogonal
polarizations, moving in opposite directions. 
The lasers
are tuned 6.6\,GHz below resonance to avoid Rayleigh scattering.
The laser intensities (about 2 mW/cm$^2$) are adjusted to keep the
diffraction efficiency below 20\,\%.

\begin{figure}[!t]
\includegraphics[width=13 cm]{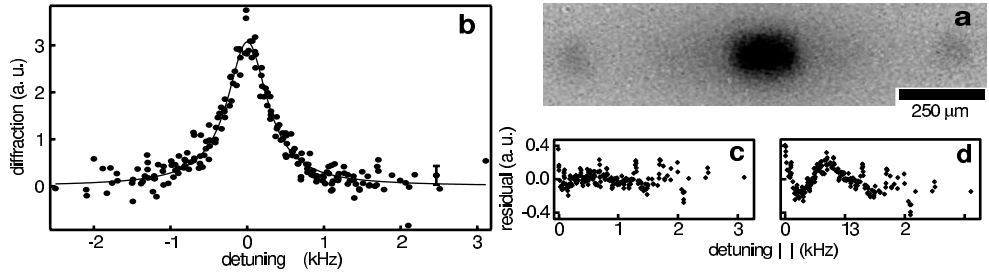}
\caption{{\bf (a)} Absorption image of degenerate cloud (center)
and diffracted atoms (left and right), averaged over several
shots, after free-flight expansion. {\bf (b)} Diffraction
efficiency (rescaled) versus relative detuning of the Bragg lasers
at $T=261(13)$\,nK, corresponding to $T/T_\phi=20(2)$. A typical
statistical error bar is shown. This spectrum is the superposition
of 12 ``elementary spectra'', as described in the text. The true
average center is 30.18(2)\,kHz, close to 30.17\,kHz, the
four-photon recoil frequency. The solid line is a Lorentzian fit,
giving a half-width at half-maximum (HWHM) of 316(10)\,Hz. {\bf
(c)} and {\bf (d)} show respectively the residual of a Lorentzian
and of a Gaussian fit to the above spectrum. Residuals are folded
around the $\delta=0$ axis, and smoothed with a six-point-wide
sliding average. } \label{spectres}
\end{figure}
Bragg spectra have been taken at various temperatures between
90(10)\,nK and 350(20)\,nK, while $T_{\rm c}$ varied from
280(15)\,nK to 380(20)\,nK. The temperature was fixed to within
20\,nK by controlling the final trap depth to a precision of
2\,kHz.
The line shape of the %resulting 
spectra is clearly
Lorentzian, not Gaussian (see Fig.~\ref{spectres}). This is a
significant result, because a Lorentzian-like profile is expected
for a momentum distribution dominated by phase fluctuations (see
\cite{Fabrice_theory}), in contrast to the gaussian-like
profile expected for a pure condensate \cite{zambelli00,stenger99}.
From the Lorentzian fit, we extract the measured half-width
$\Delta\nu_\textrm{M}$ for each temperature.

\begin{figure}[t!]
\includegraphics[width=10cm]{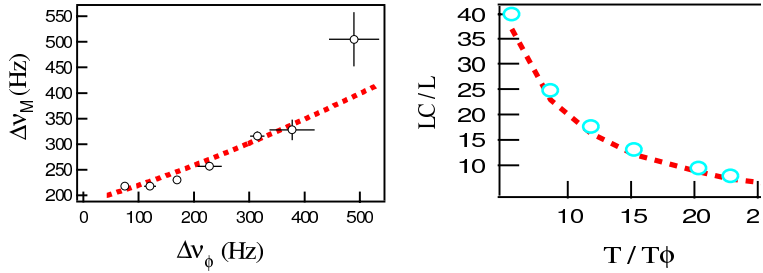}
\caption{Left : Half-widths at half-maximum $\Delta\nu_{\rm M}$ of the
experimental Bragg spectra versus the parameter
$\Delta\nu_\phi\propto \hbar / L_\phi$ (see text). Vertical error
bars are the standard deviations of the fit width; the horizontal
error bars are the one-sigma statistical dispersions of
$\Delta\nu_\phi$. The solid line is a fit assuming a Voigt profile
for the spectra. Right : coherence length deduced from the width, after correction for the gaussian broadening,} \label{largeur_vs_temp}
\end{figure}
Figure \ref{largeur_vs_temp} shows the measured spectral width $\Delta\nu_\textrm{M}$ versus
$\Delta \nu_\phi$. Dispersion of the measured $n_1(0)$ and $T$
results in the horizontal error bars on $\Delta\nu_\phi$, while
vertical error bars indicate the standard deviation of the fit
width. We assume that all experimental broadenings result
in a Gaussian apparatus function of half-width $w_{\rm G}$, to be
convolved by the Lorentzian momentum profile with a half-width
$\alpha\Delta\nu_\phi$. The convolution, a Voigt profile, has a
half-width $\alpha\Delta\nu_\phi/2+\sqrt{w_{\rm
G}^2+(\alpha\Delta\nu_\phi)^2/4}$. Note that fitting a Voigt
profile instead of a Lorentzian to a spectrum gives the same total
HWHM to less than 5\,\%, but the Lorentzian shape is too
predominant to extract reliably the Gaussian and the Lorentzian
contributions to the profile. Using $\alpha$ and $w_{\rm G}$ as
free parameters to fit the data of Fig.~\ref{largeur_vs_temp}, we
find $w_{\rm G}=176(6)$\,Hz, and $\alpha= 0.64(5)(5)$. The first
uncertainty quoted for $\alpha$ is the standard deviation of the
fit value. The second results from calibration uncertainties on
the magnification of the imaging system and on the total atom
number, which do not affect $w_{\rm G}$. The agreement of the
measured value of $\alpha$ with the theoretical value 0.67, to
within the 15\,\% experimental uncertainty, confirms
quantitatively the temperature dependence of the momentum width
predicted in Ref.~\cite{petrov3d}. The coherence length
$\hbar/p_\phi$ deduced from this measurement varies between
$5.9(8)$ and $39(4)\,\mu$m, in the range $6<T/T_\phi<28$.

\subsection{Interferometry with elongated BECs}
\label{interf}

\begin{figure}[t!]
\hskip1cm
\includegraphics[width=4.5cm]{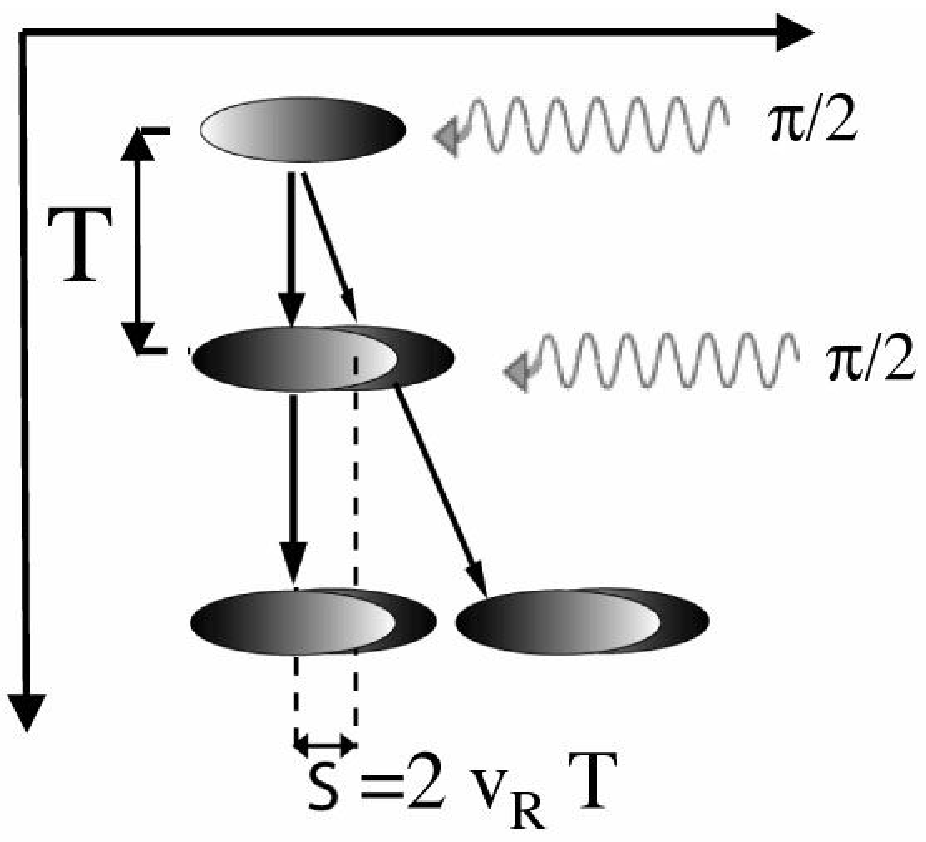}
\hskip.5cm
\includegraphics[width=4.5cm]{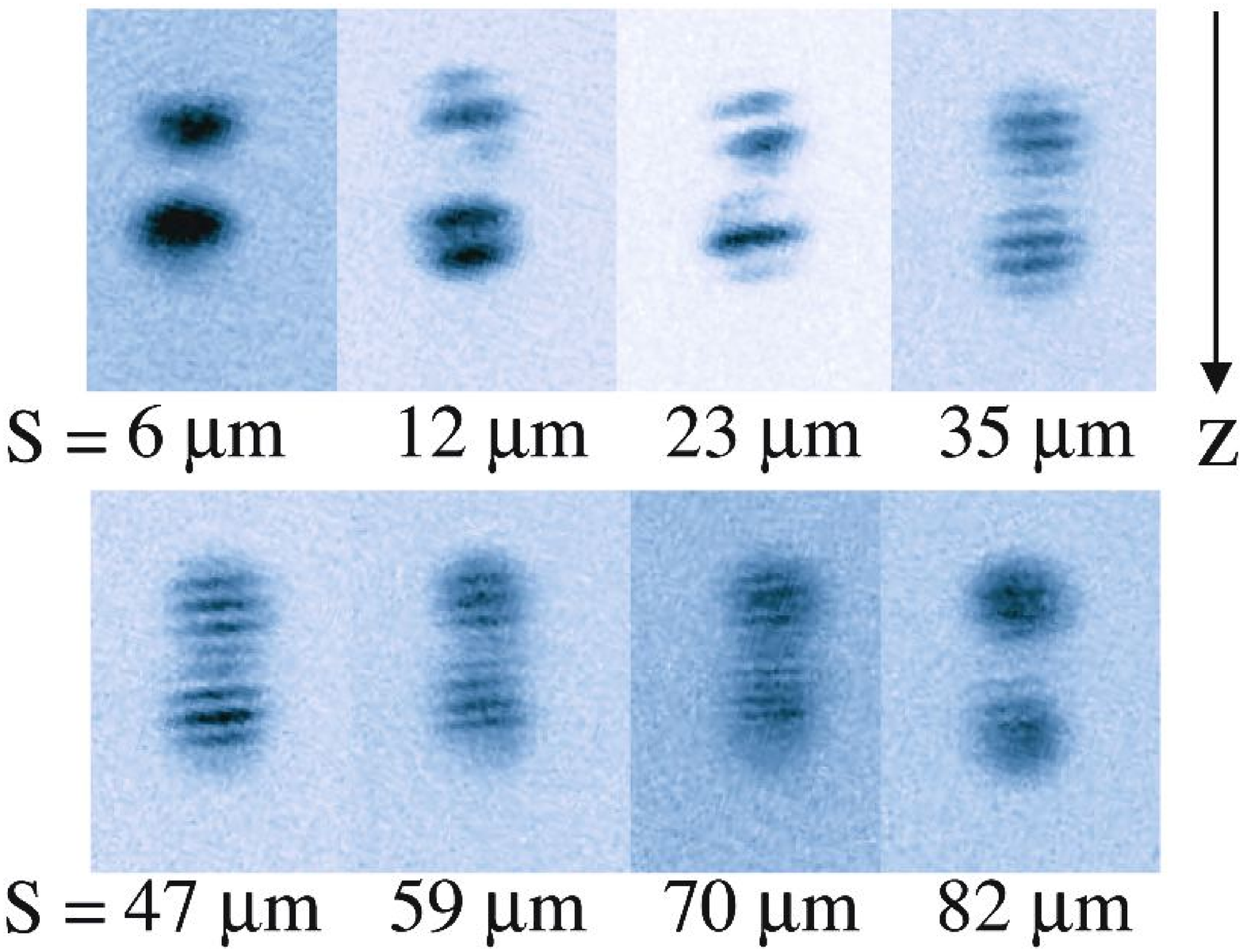}
\caption{Left : The atom interferometer used for our coherence length measurement uses 2 $\pi/2$ pulse to create and overlap 2 displaced copies of the same condensate. Right : The fringe spacing and contrast changes as the time between the 2 pulses (i.e. the separation between the 2 copies) is changed. A measurement of the fringe contrast directly measures the spatial first-order correlation function of the matter-wave field.} 
\label{interf_graph}
\end{figure}

\begin{figure}[b!]
\hskip1cm
\includegraphics[width=5.5cm]{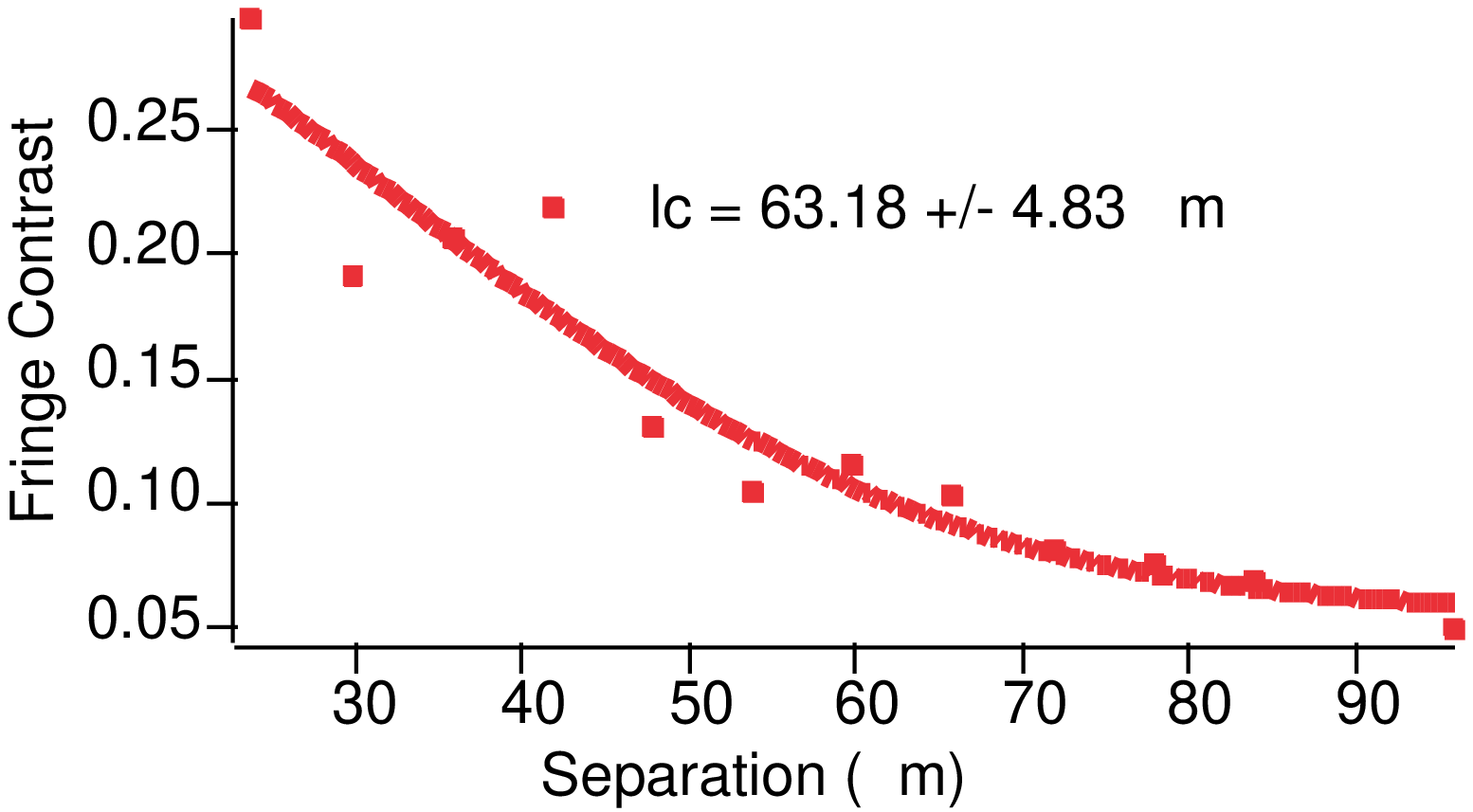}
\hskip.5cm
\includegraphics[width=5.5cm]{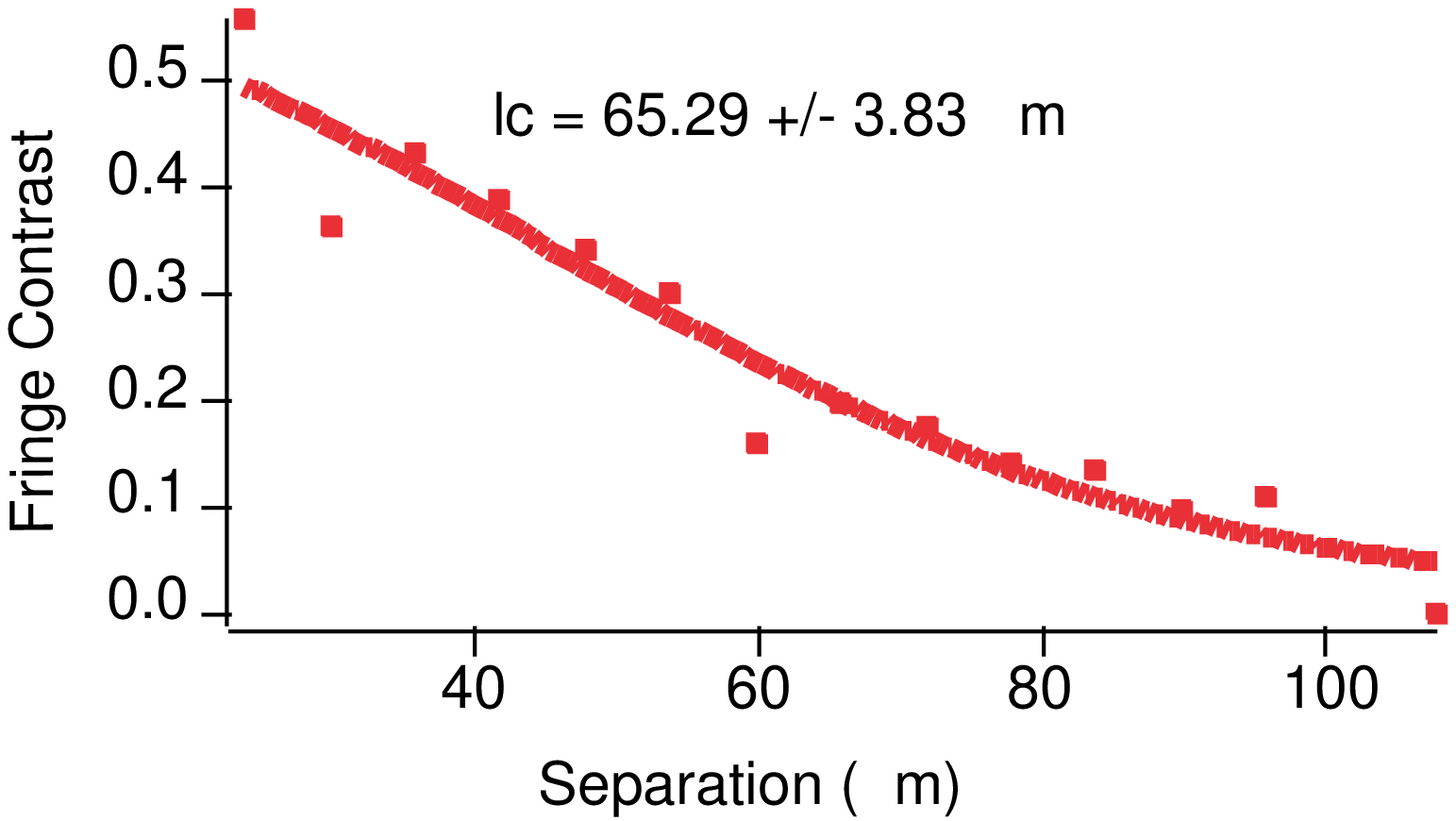}
\caption{Example of 2 direct measurements of the correlation function (fringe contrast) as a function of the separation $s$ between the two copies of the condensate ; left : $T_\Phi/T=0.45$ and $2\times 10^5$ atoms ($L_{\rm TF}\sim 100 \mu$m), right : $T_\phi/T=0.75$ and $1.3\times  10^5$ atoms ($L_{\rm TF}\sim 90 \mu$m)} 
\label{interf_result}
\end{figure}

The Bragg spectroscopy method has proven to be appropriate for measuring short coherence lengths. However, when applied to nearly coherent condensates ($L_C\sim
L$), the extremely narrow Bragg resonance becomes increasingly
difficult to resolve (for $L_\Phi\sim 40 \mu$m, the momentum width is already of the order of our resolution).
However, since our goal is to investigate the cross-over from a phase coherent to a phase-fluctuating condensate, we need to measure coherence length larger than the $39(4)\,\mu$m attainable with Bragg spectroscopy.
For this, we have implemented a matter-wave interferometer \cite{Ertmer03}. As shown on figure \ref{interf_graph}, this interferometer uses a first $\frac{\pi}{2}$ bragg pulse to split the condensate in 2 equal copies that fly apart from each other at a velocity $2v_R=2\frac{\hbar k}{M}\simeq 1.2$ cm/s.
After a time of flight $2 {\rm ms} < T < 10 {\rm ms}$, the 2 copies, that separate by $2v_R\times T$, are recombined using a second $\frac{\pi}{2}$ pulse, that creates 2 interferometer output (see fig. \ref{interf_graph}). For finite T, complementary fringes appear in
the two output ports, because of the smoothly varying expansion
phase that results from the conversion of a small fraction of the
initial mean-field energy ($\sim \epsilon^2 \mu$) into axial
kinetic energy. The resulting fringe pattern is Fourier analyzed
to find the contrast and the fringe spacing in reciprocal space.
In this way, we filter out shot-to-shot, global phase fluctuations
that would result in a translation of the fringe pattern as a
whole. This method allows for averaging many processed images
without artificially reducing the contrast. After this processing,
this method allows a direct measurement of the correlation
function $\mathcal{C}^{(1)}$ as a function of separation $s$ (see figure \ref{interf_result}),
irrespective of the amount of phase fluctuations.

\section{Conclusion}

We have demonstrated two important features of
elongated Bose-Einstein condensates: (i) the dimension of the condensate after expansion; (ii) the momentum distribution
of quasicondensates; Our method, along with
 atom interferometry can be  applied to
investigate how long range order develops during the condensate
growth.

%%-----------------------------
%%      your bibliography
%%-----------------------------


\begin{thebibliography}{99}
\bibitem{calib_N0}{The atom number measurement is calibrated to a 20\,\% precision
by comparing a measurement of $N_{\rm c}$ to theory
\cite{DalfovoRMP}.}
\bibitem{TFlimits}{Using the relevant expressions for $n_{1}(0)$, the results obtained in \cite{petrov1d} for the 1D case and
\cite{petrov3d} for the 3D case are recovered, in the limit
$L_{\phi} \ll L$.}
\end{thebibliography}
\end{document}